\documentclass[%
reprint,
aps,
prl,
amsmath,
twocolumn,
amssymb,
floatfixng,
showpacs,
superscriptaddress,
footinbib,
multicol
]{revtex4-2}
\usepackage{graphicx}
\usepackage{epstopdf}
\usepackage{comment}
\usepackage[justification=justified]{caption}
\usepackage{dcolumn}
\usepackage{bm}
\usepackage{color}
\usepackage{xcolor}
\usepackage{amsmath}
\usepackage{braket}
\usepackage{amsfonts}
\usepackage{soul}
\usepackage{url}
\usepackage{array,tabularx}
\usepackage{graphics}
\usepackage{graphicx}
\usepackage{subcaption}
\usepackage{times}

\usepackage{caption}
\DeclareCaptionJustification{fulljustified}{\leftskip=0pt\rightskip=0pt\parfillskip=0pt plus 1fil}
\usepackage[export]{adjustbox}
\usepackage[colorlinks=true,linktoc=page,linkcolor=red,citecolor=blue]{hyperref}

\usepackage{xurl}      %
\usepackage{hyperref}  %

\definecolor{OliveGreen}{rgb}{0,0.6,0}

\begin{document}

\title{
Acoustic Illusion of a Rigid Object for a Moving Emitter via Equivalent Surface Control}

\author{Asaf Kor}
\affiliation{School of Mechanical Engineering, Tel Aviv University, Tel Aviv 69978, Israel}

\author{Lea Beilkin-Sirota}\email{leabeilkin@tauex.tau.ac.il}
\affiliation{School of Mechanical Engineering, Tel Aviv University, Tel Aviv 69978, Israel}

\begin{abstract}
We report a three-dimensional acoustic illusion of a rigid object illuminated by an emitter moving along an arbitrary trajectory. The object is absent: control sources are arranged around its location, the closed surface they define being termed the equivalent surface. Driven by the field the object would have scattered onto this surface, the sources radiate, by the Kirchhoff--Helmholtz representation, that same field beyond it. The control signals follow in closed form from a time-dependent physical optics formulation that we derive for a moving emitter, in which the motion enters analytically through the acoustic travel times, so that the scattering problem need not be re-solved at each emitter position, nor the propagation domain discretized. For objects too large to enclose, the signals are mapped onto an internal surface, an option whose advantages and penalties are discussed. Numerical simulations validate the method in both configurations.
\end{abstract}

\maketitle

\section{Introduction} \label{sec:intro}

Acoustic illusion refers to the synthesis of a prescribed acoustic field such that a receiver array or imaging system perceives an object that is not physically present. The acoustic signature to be reproduced may be either the echoes a target object scatters when probed by an external sound or the noise it radiates of its own accord. 
The applications are broad. In nondestructive testing, defect signatures can be emulated for the calibration of inspection procedures~\cite{Virkkunen2021,Mohseni2021}. In underwater acoustics, entire scenes can be rendered for sonar calibration and operator training, reducing dependence on costly at-sea trials~\cite{Foote2005,Jung2013}. The defense counterpart is sonar deception, in which an adversary's sonar is presented with the response of a target of chosen geometry or motion~\cite{Chen2018}.

Active illusions have been pursued by several approaches. In three-dimensional wavefield engineering, control sources are usually placed at arbitrary positions, and their driving signals are obtained by numerical optimization, solving an inverse problem that matches a prescribed target field at a set of control points. The inversion is typically performed in the frequency domain, using regularized least squares with measured or analytical transfer functions~\cite{kirkeby1993reproduction,kim2004optimal,cheer2016active}, and underlies applications ranging from personal audio, sound zoning, and cross-talk cancellation, to stealth and large-scale outdoor sound field control~\cite{olivieri2016beamforming,koyama2020optimizing,fazi2025benefits,eggler2019active,heuchel2020outdoor}. In one or two dimensions, physics-based control may bypass numerical minimization, producing illusions of spatial or spatio-temporal voids~\cite{lasri2023real,lasri2023active}.

Another approach is based on integral representations, specifically the Kirchhoff--Helmholtz (KH) theorem~\cite{williams1999fourier,pierce2019acoustics}. Following from Green's second identity, the theorem states that if the acoustic pressure and its normal derivative are known on a closed surface, the field is determined at every point of the source-free region, while the integral vanishes in the complementary one. The boundary quantities may be time varying, the representation then holding in the time domain. 

\begin{figure*}[!tb]    \centering
\begin{tabular}{cc}
 \includegraphics[width=1\columnwidth, valign = c]{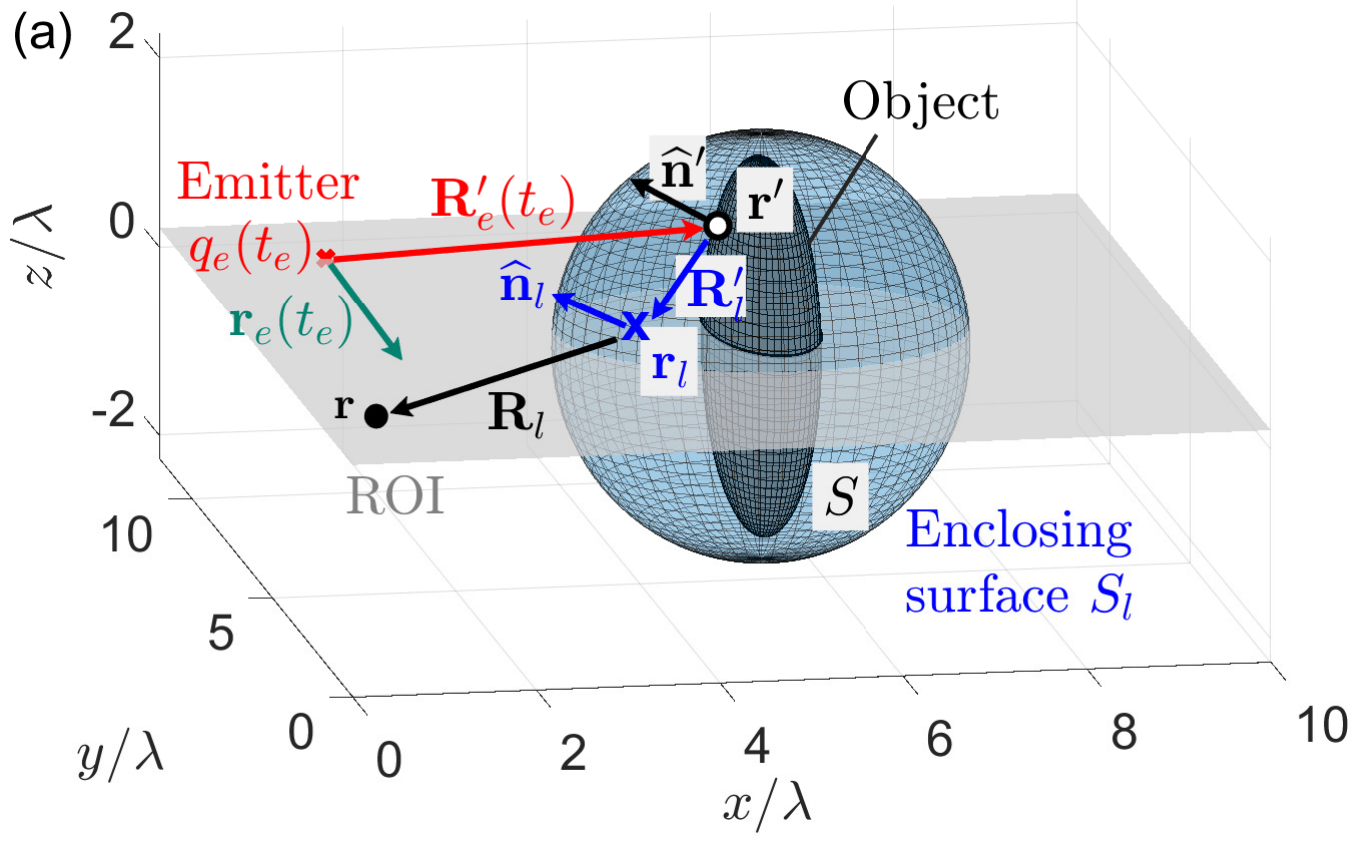}    &  \includegraphics[width=1\columnwidth, valign = c]{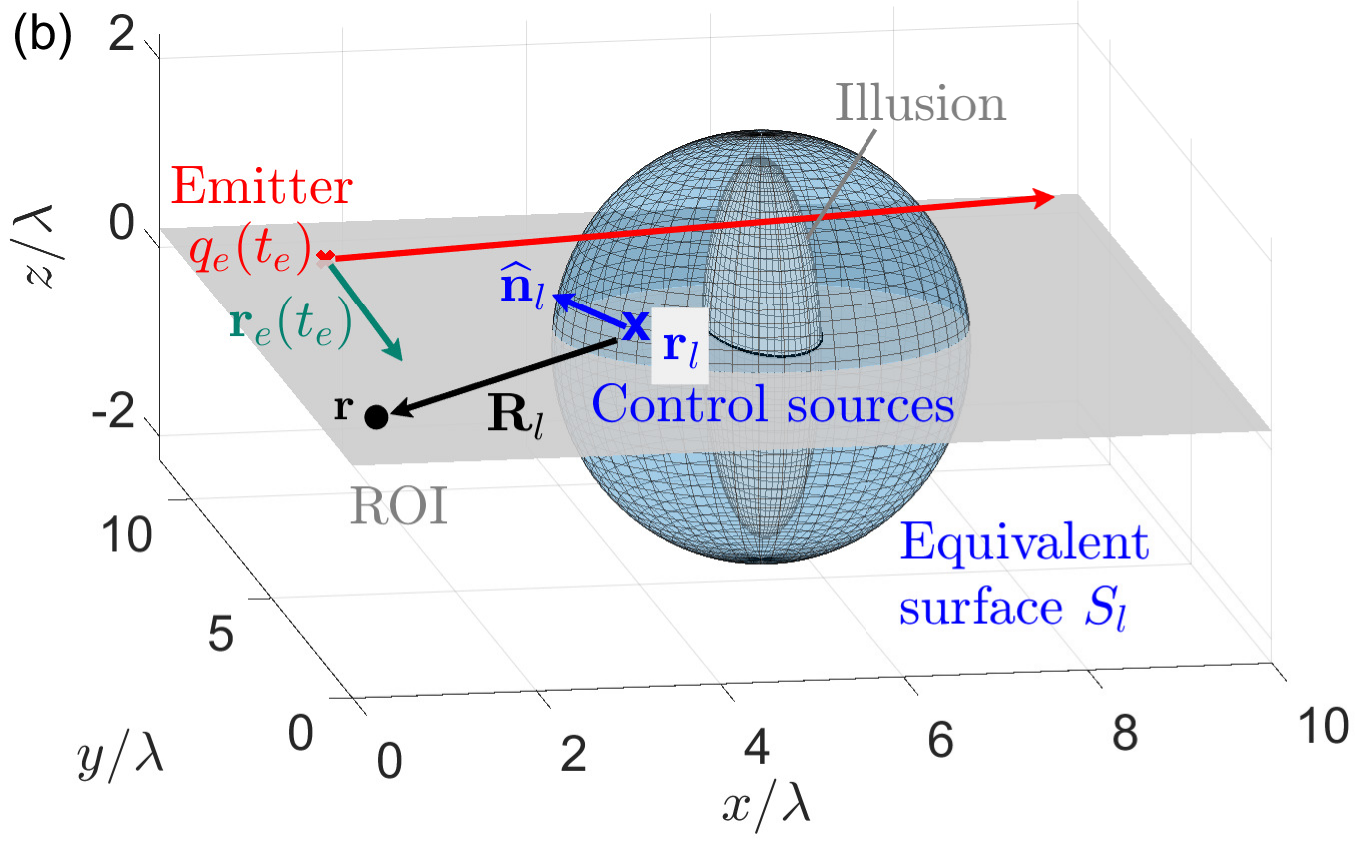}
\end{tabular}
    \caption{Model schematics. (a)~Kirchhoff--Helmholtz representation. An emitter
$q_e$ outside the enclosing surface $S_l$ illuminates a rigid object, and the
scattered field at any exterior point $\mathbf{r}$ follows from the
scattered pressure and normal velocity at all points $\mathbf{r}_l$ on $S_l$. (b)~Illusion based on the
representation. The object is absent and $S_l$ is an equivalent surface
carrying monopole and dipole control sources driven by these quantities, reproducing the scattered
field in the region of interest (ROI).}
    \label{fig:model_intro}
\end{figure*}

Two branches follow, according to which side the sources lie on. The interior form, with sources outside the surface and the field reconstructed within, underlies Huygens' principle. In the exterior form, adopted here and illustrated in Fig.~\ref{fig:model_intro}(a), the sources are enclosed and the field is reconstructed outside, the radiation condition ensuring that the contribution from the surface at infinity vanishes.

The illusion follows by reading the theorem in reverse. Rather than evaluating a field from known boundary data, the scattered field that the emulated object would produce is computed on a chosen enclosing surface and used to drive control sources distributed over it, as illustrated in Fig.~\ref{fig:model_intro}(b). By the exterior form of the theorem, the array then radiates precisely the field the object would have scattered, although the object is absent. The surface is acoustically transparent and exists only as the locus of the sources; it is therefore termed an equivalent surface, or control surface, and its shape is unconstrained provided it is closed and encloses the object.

This approach underlies the immersive boundary condition framework~\cite{vanmanen2015broadband}, in which sources on a closed surface are driven so as to reproduce the response of a virtual scatterer. The transfer functions relating the surface sources to the field are obtained either from full-waveform finite-element simulations of the emulated object~\cite{becker2021broadband}, or retrieved directly from measurements of a physical specimen~\cite{muller2023cloning}. Because these experiments are performed in a laboratory of finite extent, the emulation is accompanied by a second task, namely accounting for the reflections of enclosing boundaries.

Here we address an emitter in motion in an unbounded three-dimensional medium. Its signal and position are taken as available while it moves, the trajectory being either prescribed in advance or estimated in real time by an independent tracking algorithm.
In principle, the control signals may be evaluated by any method capable of predicting the scattered field of the object. The finite element method (FEM)~\cite{ihlenburg1998fem} and the finite-difference time-domain (FDTD) method~\cite{botteldooren1995fdtd} discretize the propagation domain and require truncation by absorbing boundaries, while the boundary element method (BEM)~\cite{wu2000bem} discretizes only the object surface, but couples every surface element to every other, so that a dense linear system must be solved.

We adopt instead physical optics (PO), which approximates the surface field directly, taking it to be that of a locally flat rigid reflector over the illuminated portion and zero elsewhere, so that the scattered field follows by direct integration~\cite{MorseIngard1968,ColtonKress2019,Balanis2012,ufimtsev2014fundamentals}. Shadowing and edge diffraction are thereby reproduced approximately, the shadow forming where the scattered field cancels the incident one. No system is solved and no propagation domain is meshed. The technique works best for objects with local radii of curvature large compared to the wavelength.

Since the problem considered here is transient, owing both to the emitter's motion and to the possible transiency of its signal, the established frequency-domain form of PO cannot be applied directly. A time-domain version of the method is therefore required. We recently derived such a formulation for a moving emitter~\cite{kor2026tdpo}, referred to here as time-dependent physical optics (TDPO), whose main principles are recapped in Sec.~\ref{sec:time_domain_regular}.

Moving sources can also be accommodated in time-domain FEM, by redistributing the source forcing among the nodes as it advances or updating the mesh~\cite{Manduzio2021}, in BEM, by evaluating the incident field on the object boundary at the retarded emission times~\cite{AntesMeise1991}, and in FDTD, by interpolating the source onto the fixed grid~\cite{TsuchiyaKanamori2021}. These methods, however, still operate on a discretized domain or boundary, whereas in TDPO the scattered field follows by direct integration, at a far lower computational cost.

Combining TDPO with the time-domain KH representation, the control signals on the equivalent surface follow in closed form for an emitter moving along an arbitrary trajectory, with no discretization of the propagation domain. We first develop and validate this approach in the main configuration, in which the equivalent surface encloses the emulated object (Sec.~\ref{sec:regular}). As enclosure may become impractical for large objects, we then consider a smaller equivalent surface that does not enclose it (Sec.~\ref{sec:redatumed}), and discuss the benefits and limitations of the two configurations (Sec.~\ref{sec:conclusion}).

\section{Illusion via an enclosing control surface}  \label{sec:regular}

The setup is defined in Fig.~\ref{fig:model_intro}. The emulated object
is perfectly rigid, stationary, and defined by the bounding surface $S$.
The formulation applies to an arbitrary object geometry; a simple convex shape
is depicted in Fig.~\ref{fig:model_intro} for clarity, whereas the numerical
examples below use a more general one. 
The surrounding medium is
homogeneous, isotropic and lossless, with mass density $\rho$ and
sound speed $c$, and is unbounded.

The illuminating emitter is a monopole of strength $q_e(t_e)$ (in
$\mathrm{kg/s^2}$), the time derivative of its mass injection rate, following a trajectory $\mathbf{r}_e(t_e)$, where $t_e$
denotes the emission time. In the absence of the object it produces the
incident field $p_{\mathrm{inc}}(\mathbf{r},t)$ at an observation point
$\mathbf{r}$ and observation time $t$; were the object present, the incident
field would induce a scattered field $p_{\mathrm{scat}}(\mathbf{r},t)$, and the
total field would be $p_{\mathrm{tot}} = p_{\mathrm{inc}} + p_{\mathrm{scat}}$.
The region of interest (ROI), on which the illusion is to be generated, may be
any subset of the three-dimensional space; for clarity of presentation it is
taken here as the plane $z = 0$, with the emitter moving within it.

The control sources are distributed over the equivalent surface $S_l$, assumed
closed and sufficiently smooth but otherwise arbitrary, provided that $S$ lies
inside the volume it bounds; a spherical $S_l$ is used in Fig.~\ref{fig:model_intro}. They
are monopoles and dipoles of strengths $q$ and $F$, the dipoles
oriented along the outward normal $\widehat{\mathbf{n}}_l$. The theorem
requires both layers on $S_l$, but not necessarily at the same points; they are
collocated here for convenience. The continuous
description is an idealization: in practice the sources form a discrete array,
sampled according to the wavelength $\lambda$. 

The task is to determine the control signals $q$ and $F$ such that
the field radiated by the control sources reproduces $p_{\mathrm{scat}}$
throughout the ROI. We first
consider the established case of a stationary emitter radiating a harmonic
signal, for which a steady state is reached and the problem admits a
frequency-domain formulation, with the scattered field evaluated by PO (Sec.~\ref{sec:freq}). The time-domain formulation for a moving emitter follows in Sec.~\ref{sec:time_domain_regular}.

\subsection{Frequency domain formulation}  \label{sec:freq}

In this section the emitter is stationary, $\mathbf{r}_e(t_e) = \mathbf{r}_e$,
and radiates a harmonic signal of complex strength $\widetilde{q}_e(\omega)$ at angular frequency $\omega$, so
that a steady state is reached. We assume
time-harmonic dependence $e^{\mathrm{j}\omega t}$, suppressed throughout for
brevity. Were the object present, as in Fig.~\ref{fig:model_intro}(a), its
scattered field on $S_l$ would have pressure $\widetilde{p}_{\textrm{scat}}$ and
outward normal velocity
$\widetilde{v}_{\textrm{scat}} = \widetilde{\mathbf{v}}_{\textrm{scat}}\cdot
\widehat{\mathbf{n}}_l$, the latter scaled as $\widetilde{u}_{\textrm{scat}} =
\mathrm{j}\omega\rho\,\widetilde{v}_{\textrm{scat}}$, so that $\widetilde{u}_{\textrm{scat}} =
-\partial_{n_l}\widetilde{p}_{\textrm{scat}}$. With
$\mathbf{R}_l = \mathbf{r} - \mathbf{r}_l$ the vector from a control source to
the observation point and $R_l = |\mathbf{R}_l|$, the scattered field at any
$\mathbf{r}$ exterior to $S_l$ follows from the KH representation as
\begin{equation}  \label{eq:holography_S_l}
\begin{aligned}
{\widetilde{p}}_{\textrm{scat}}(\mathbf{r};\omega)
&=\int_{S_l} \Big[G^{p|q}(R_l;\omega)\,\widetilde{u}_{\textrm{scat}}(\mathbf{r}_l;\omega) \\
&\qquad + G^{p|F}(R_l;\omega)\,\widetilde{p}_{\textrm{scat}}(\mathbf{r}_l;\omega)\Big]\,\mathrm{d}s_l ,
\end{aligned}
\end{equation}
the same integral vanishing for $\mathbf{r}$ interior to $S_l$.

In the illusion, Fig.~\ref{fig:model_intro}(b), the monopole and dipole layers of control sources on $S_l$ are thus driven, respectively, by  $\widetilde{u}_{\textrm{scat}}(\mathbf{r}_l;\omega)$ and  $\widetilde{p}_{\textrm{scat}}(\mathbf{r}_l;\omega)$. Since \eqref{eq:holography_S_l} integrates over $S_l$, both are surface densities, in $\mathrm{kg/(m^2\,s^2)}$ and $\mathrm{Pa}$. A realizable array is discrete, its $L$ sources at $\mathbf{r}_l^{(i)}$ carrying the lumped strengths
\begin{subequations} \label{eq:lumped}
\begin{align}
\widetilde{q}^{(i)}(\omega)&=\widetilde{u}_{\textrm{scat}}(\mathbf{r}_l^{(i)};\omega)\,\mathrm{d}s_l^{(i)}, \label{eq:lumped_q}\\
\widetilde{F}^{(i)}(\omega)&=\widetilde{p}_{\textrm{scat}}(\mathbf{r}_l^{(i)};\omega)\,\mathrm{d}s_l^{(i)}, \label{eq:lumped_F}
\end{align}
\end{subequations}
in $\mathrm{kg/s^2}$ and $\mathrm{N}$, respectively. The monopole is a point volume source, $\widetilde{q}^{(i)}$ being $\mathrm{j}\omega\rho$ times its volume velocity $\widetilde{Q}^{(i)}(\omega)$, and the dipole is a point force $\widetilde{\mathbf{F}}^{(i)}=\widetilde{F}^{(i)}\,\widehat{\mathbf{n}}_l$ directed along the outward normal. The field radiated by the discrete array, the controlled counterpart of \eqref{eq:holography_S_l}, then approximates the scattered field as
$\widetilde{p}_{\textrm{scat}}(\mathbf{r};\omega)
\approx \sum_{i=1}^{L}[G^{p|q}(R_l^{(i)};\omega)\,\widetilde{q}^{(i)}(\omega)
+ G^{p|F}(R_l^{(i)};\omega)\,\widetilde{F}^{(i)}(\omega)]$.
Here, $G^{p|q}$ and $G^{p|F}$ denote the corresponding Green’s function kernels, given by
\begin{subequations} \label{eq:Gp}
    \begin{align}
        G^{p|q}(R_l;\omega)&=G(R_l;\omega), \label{eq:GpQ}\\        
        G^{p|F}(R_l;\omega) &= G(R_l;\omega)\,\alpha_l\,
                       \widehat{\textbf{R}}_l\cdot\widehat{\textbf{n}}_l,  \label{eq:GpF}
    \end{align}
\end{subequations}
where $\alpha_l=\mathrm{j}k+1/R_l$, with $k=\omega/c$ the wavenumber, $G^{p|F}=\partial_{n_l} G = \nabla_{\mathbf r_l} G \cdot \hat{\mathbf n}_l$, and $\widehat{\mathbf{R}}_l=\mathbf{R}_l/R_l$. The function $G(r;\omega)$ is the three-dimensional free-space Green's function, which for a general source--observer distance $r$ is defined as~\cite{pierce2019acoustics}
\begin{equation}  \label{eq:Green_q}
    G(r;\omega)=\frac{e^{-\mathrm{j}kr}}{4\pi r}.
\end{equation}
\begin{figure*}[tb]    
\centering
\renewcommand{\arraystretch}{2} 
\begin{tabular}{c}
\begin{tabular}{ccc}
  \includegraphics[height=4.3 cm, valign = c]{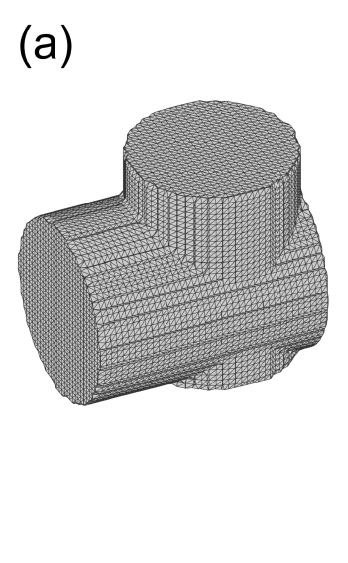}   &  \includegraphics[height=4.3 cm, valign = c]{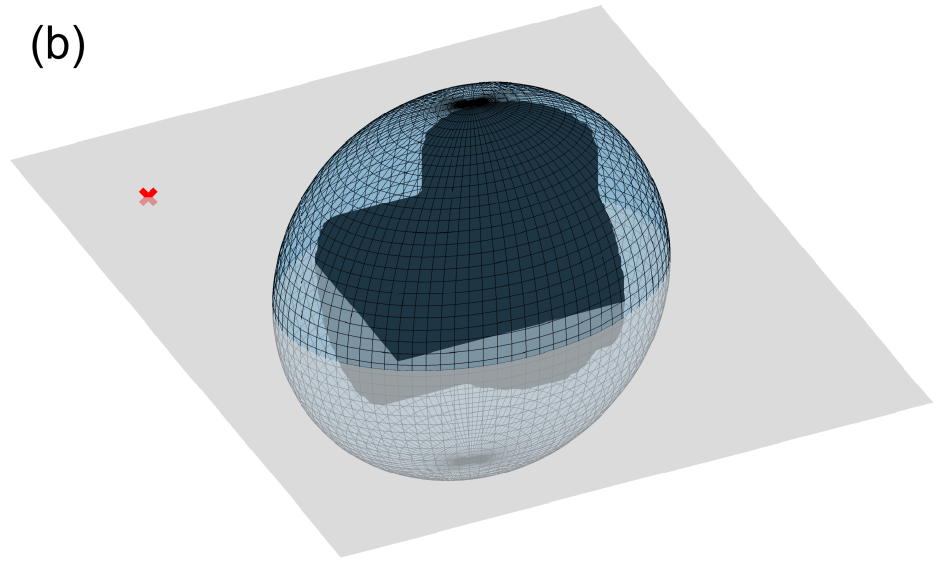} & \includegraphics[height=4.3 cm, valign = c]{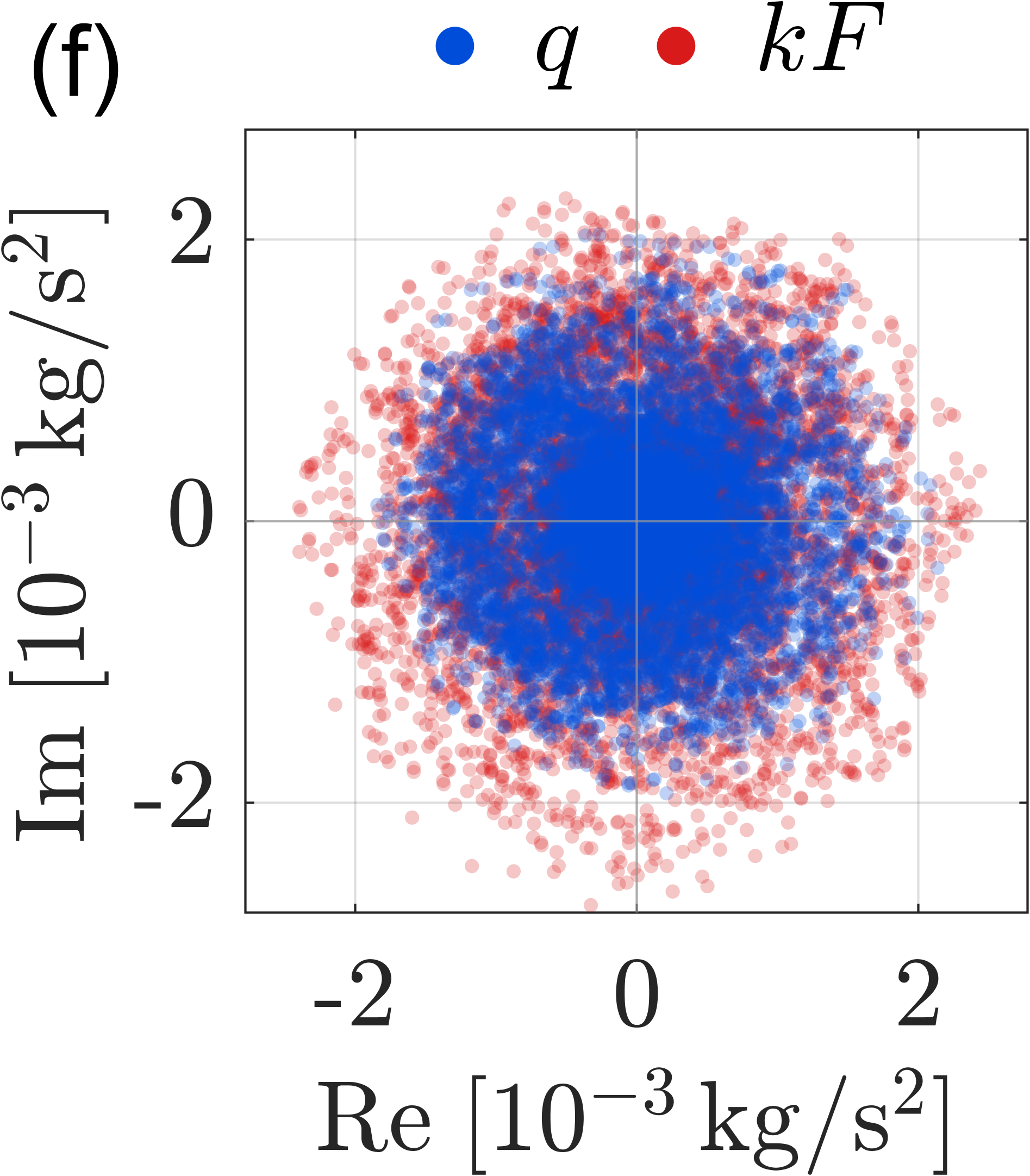}
\end{tabular}
 \\ \\
\setlength{\tabcolsep}{0pt} 
\begin{tabular}{ccc}
 \includegraphics[height=5.1 cm, valign = c]{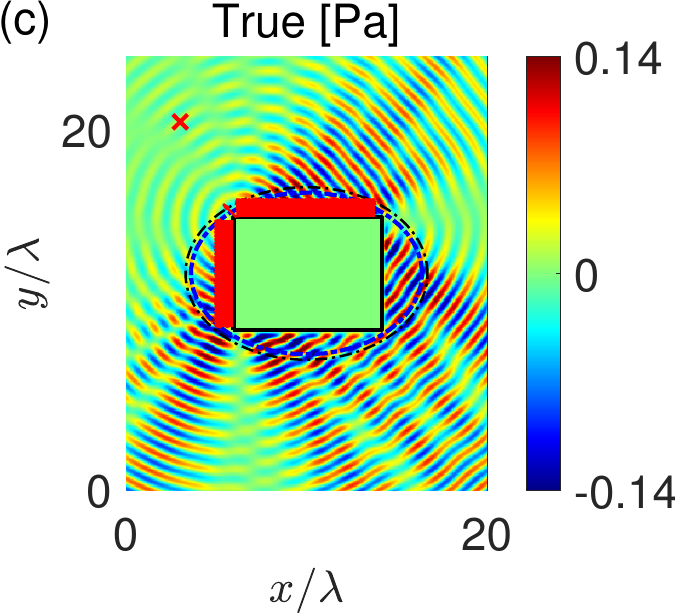}    &  \includegraphics[height=5.1 cm, valign = c]{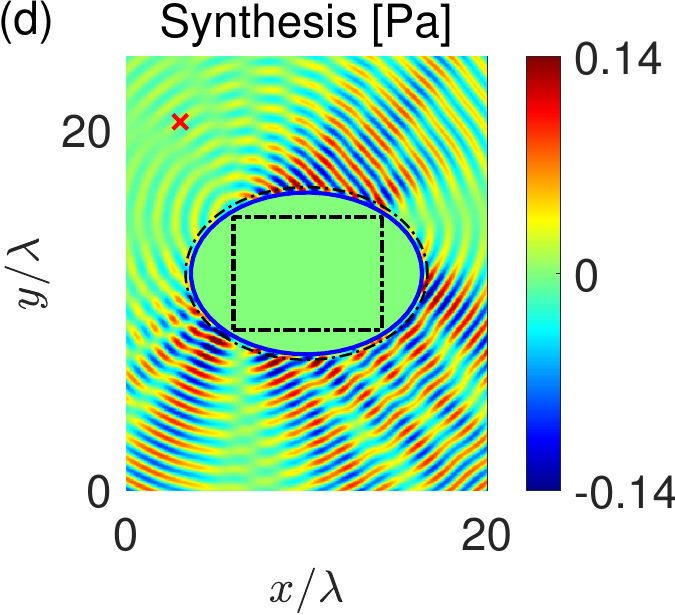}
     &  \includegraphics[height=5.1 cm, valign = c]{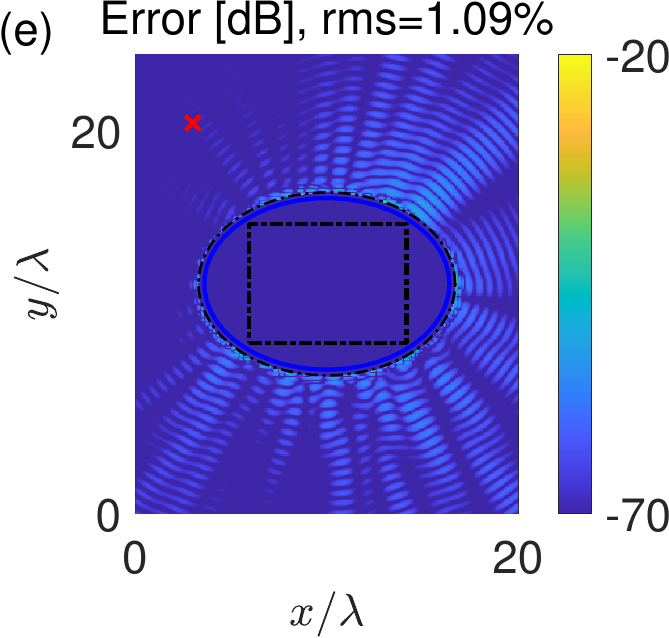}
\end{tabular}
\end{tabular}
\caption{Validation of the control algorithm in the frequency domain via
numerical simulations based on \eqref{eq:holography_S_l}--\eqref{eq:vectors_obj}.
(a)~The object. (b)~The ellipsoidal equivalent surface $S_l$, with the object,
the ROI plane and the emitter (red cross). (c)--(e)~Top views of the ROI: the
true scattered field, the synthesized illusion field and their difference in
dB re 1~Pa, with the rms error in the title of (e). The black rectangle is the cross section of
the object, with the outward normals at its illuminated points in red, and the
blue curve is the intersection of $S_l$ with the ROI; the field is not evaluated
inside them and is shown as zero. (f)~The control signals at all sources, in the
complex plane.}
    \label{fig:freq_sim}
\end{figure*}

The kernels $G^{p|q}$ and $G^{p|F}$ thus account for propagation from $S_l$ to
the ROI, so that the reconstruction \eqref{eq:holography_S_l} follows once the
control signals \eqref{eq:lumped} are determined. As stated in Sec.~\ref{sec:intro}, these follow from $\widetilde{u}_{\textrm{scat}}(\mathbf{r}_l;\omega)$ and
$\widetilde{p}_{\textrm{scat}}(\mathbf{r}_l;\omega)$, which can in principle be
evaluated by BEM or by grid-based methods such as FEM and FDTD, and are
obtained here from the PO approximation.

By the PO approach, the scattered pressure at a point $\mathbf{r}_l$ on $S_l$ is given by the contribution of the lit portion of the object surface $S$, defined by $\mathbf{R}_e' \cdot \widehat{\mathbf{n}}' < 0$, as \cite{MorseIngard1968,ColtonKress2019,Balanis2012,ufimtsev2014fundamentals}
\begin{equation}  \label{eq:p_scat_l}
    \widetilde{p}_{\textrm{scat}}(\mathbf{r}_l;\omega)
    = \int_{S_{\textrm{lit}}}
    2\widetilde{q}_e(\omega)\,G(R'_e;\omega)\,G(R'_l;\omega)\,
    \alpha'_l\,\widehat{\mathbf{R}}'_l\cdot \widehat{\mathbf{n}}'\,
    \mathrm{d}s'.
\end{equation}
Here, $\widehat{\mathbf{n}}'$ is the outward unit normal at the point $\mathbf{r}' \in S$, and
\begin{equation} \label{eq:vectors_obj}
\mathbf{R}'_e = \mathbf{r}' - \mathbf{r}_e, \qquad
\mathbf{R}'_l = \mathbf{r}_l - \mathbf{r}' ,
\end{equation}
are the vectors from the emitter to $S$ and from $S$ to $S_l$, with
$R'_e = |\mathbf{R}'_e|$, $R'_l = |\mathbf{R}'_l|$ and
$\widehat{\mathbf{R}}'_l = \mathbf{R}'_l/R'_l$.
Both are time-invariant here, the emitter, the object and $S_l$ all being
stationary. Further, $\alpha'_l$ is obtained by substituting
$R_l\rightarrow R'_l$ in $\alpha_l$ in \eqref{eq:GpF}, and $\mathrm{d}s'$ is
the differential surface element. The Green's functions $G(R'_e;\omega)$ and
$G(R'_l;\omega)$ follow from \eqref{eq:Green_q}.
For numerical evaluation, $S$ is discretized into elemental cells and the
integral over $S_{\textrm{lit}}$ is approximated by a summation. Then, $\widetilde{u}_{\textrm{scat}}(\mathbf{r}_l;\omega)$ can be
obtained analytically by differentiating the integrand of \eqref{eq:p_scat_l}
with respect to $\mathbf{r}_l$, which yields a lengthier kernel. It is more conveniently evaluated by a finite
difference between the pressures on the offset surfaces $S_l^{\pm}$, defined by
$\mathbf{r}_l \pm h\,\widehat{\mathbf{n}}_l$, each obtained from
\eqref{eq:p_scat_l}.

To demonstrate the control algorithm described by
\eqref{eq:holography_S_l}--\eqref{eq:vectors_obj}, we consider the geometrically
complex object composed of two intersecting cylinders shown in
Fig.~\ref{fig:freq_sim}(a). This object is not physically present, and the task
of the control system in Fig.~\ref{fig:freq_sim}(b) is to emulate its presence
by generating the field that would be scattered from it. The configuration
corresponds to Fig.~\ref{fig:model_intro}(b) with a stationary emitter.

The equivalent surface $S_l$ is an ellipsoid enclosing the object as tightly as
the geometry permits, so as to limit the extent of the control region, with the
sources distributed at a spacing of $\lambda/5$. The simulations are conducted
in air ($c=344$~m/s), with all spatial dimensions normalized by the wavelength, the emitter
radiating at $f=6.88$~kHz, i.e.\ $\lambda=5$~cm, with strength
$\widetilde{q}_e = 1\,\mathrm{kg/s^2}$. The two cylinders have radii of about
$3\lambda$ and $2.7\lambda$ and lengths of about $7.8\lambda$ and $7.1\lambda$,
so that $kR\approx17$--$19$ on their curved faces, well within the regime of
PO validity away from the rims and the junction. The emitter is located on
the ROI plane $z = 0$, which excludes the interior of $S_l$. The resulting
fields are presented in Fig.~\ref{fig:freq_sim}(c)--(e).

Panel (c) shows the reference scattered field on the ROI, obtained by direct PO
evaluation of the physical object using \eqref{eq:p_scat_l} with
$\mathbf{r}_l$ replaced by the observation point $\mathbf{r}$. Panel (d) shows
the reconstructed illusion field obtained from the control law
\eqref{eq:holography_S_l}--\eqref{eq:vectors_obj}, and panel (e) their
difference. The accuracy is measured by the relative root mean square (rms)
error,
$e_{\mathrm{rms}}=\big(\sum|p_{\mathrm{syn}}-p_{\mathrm{ref}}|^2/\sum|p_{\mathrm{ref}}|^2\big)^{1/2}$,
with $p_{\mathrm{ref}}$ and $p_{\mathrm{syn}}$ the reference and synthesized
fields and the sums running over the ROI points more than $\lambda/4$ outside
$S_l$; here $e_{\mathrm{rms}}=1.1\%$. Since
both fields derive from the same PO approximation, this error measures the
fidelity of the reconstruction, independently of the accuracy of PO itself.

The control signals are shown in Fig.~\ref{fig:freq_sim}(f), the dipole strengths
being scaled by $k$ so that both share the units $\mathrm{kg/s^2}$. They are
small, $|\widetilde{q}^{(i)}|$ and $|k\widetilde{F}^{(i)}|$ remaining below
about $3\times10^{-3}\,\mathrm{kg/s^2}$, some three orders of magnitude below
$\widetilde{q}_e$, since each source accounts only for the scattered field over
its patch, the $L$ contributions adding up in the ROI.

\subsection{Time domain formulation}  \label{sec:time_domain_regular}

We now let the emitter move along a trajectory $\mathbf{r}_e(t_e)$ and radiate a
possibly transient signal $q_e(t_e)$, so that no steady state is reached and the
formulation must be carried out directly in the time domain.
Defining $u_{\mathrm{scat}} = \rho\,\dot{v}_{\mathrm{scat}}$ as the counterpart of
$\widetilde{u}_{\textrm{scat}}$, the emulated scattered field at
an observation point $\mathbf{r}$ exterior to $S_l$ is expressed as a
superposition of convolutional contributions from the control surface,
\begin{equation}  \label{eq:td_KH}
p_{\mathrm{scat}}(\mathbf{r},t)
=\int_{S_l}
\left[g^{p|q}*u_{\mathrm{scat}}+g^{p|F}*p_{\mathrm{scat}}\right]
(\mathbf{r},t)\,\mathrm{d}s_l ,
\end{equation}
the time-domain counterpart of \eqref{eq:holography_S_l}. Here $*$ denotes
temporal convolution, with the arguments of the convolved quantities suppressed
for compactness. As in the harmonic case, the monopole and dipole layers are
driven by the quantities on the equivalent surface itself,
$u_{\mathrm{scat}}(\mathbf{r}_l,t_l)$ and $p_{\mathrm{scat}}(\mathbf{r}_l,t_l)$,
related to the lumped strengths of a discrete array as in Sec.~\ref{sec:freq}.
Here the retarded time $t_l=t-R_l/c$ is the arrival time of the scattered field at
$\mathbf{r}_l$, and also the emission time of the control source there.

The kernels $g^{p|q}$ and $g^{p|F}$ are the inverse transforms of
\eqref{eq:GpQ} and \eqref{eq:GpF}, built on the time-domain free-space Green's
function
\begin{equation}  \label{eq:Green_q_t}
g(r,t)=\frac{\delta(t-r/c)}{4\pi r},
\end{equation}
the counterpart of \eqref{eq:Green_q}. Since the control surface and the
observation ROI are stationary, $R_l$ is time-invariant, and with
$\mathrm{j}\omega \rightarrow \partial_{t_l}$ in $\alpha_l$ the convolutions in
\eqref{eq:td_KH} collapse onto $t_l$,
\begin{subequations} \label{eq:td_kernels}
\begin{align}
(g^{p|q}*u_{\mathrm{scat}})(\mathbf{r},t)
&=\frac{u_{\mathrm{scat}}(\mathbf{r}_l,t_l)}{4\pi R_l},
\label{eq:td_Gpq}\\
(g^{p|F}*p_{\mathrm{scat}})(\mathbf{r},t)
&=\frac{\widehat{\mathbf{R}}_l\cdot\widehat{\mathbf{n}}_l}{4\pi c R_l}
\left(\frac{\partial}{\partial t_l}+\frac{c}{R_l}\right)
p_{\mathrm{scat}}(\mathbf{r}_l,t_l).
\label{eq:td_GpF}
\end{align}
\end{subequations}
Equations \eqref{eq:td_KH}--\eqref{eq:td_kernels} describe propagation from
$S_l$ to the ROI, which in a physical implementation is carried out by the
medium itself. The control problem therefore reduces, as before, to determining
$p_{\mathrm{scat}}$ and $u_{\mathrm{scat}}$ on $S_l$.

These are obtained here from the TDPO formulation for a moving
emitter~\cite{kor2026tdpo}. A contribution emitted at time $t_e$ arrives at a
point $\mathbf{r}'$ on the emulated object boundary at $t'$, and, upon
reradiation, at a control point $\mathbf{r}_l$ at $t_l$, where
\begin{equation}  \label{eq:time_relations}
t'=t_e+\frac{R'_e(t_e)}{c},
\qquad
t_l=t'+\frac{R'_l}{c},
\end{equation}
with $R'_e$ and $R'_l$ the emitter-to-object and object-to-$S_l$ distances of
\eqref{eq:vectors_obj}. The object and $S_l$ being stationary, $R'_l$ remains
time-invariant, whereas $R'_e$ now varies with the emission time. Given $t_l$,
the second relation therefore yields $t'$ directly, while the first is implicit
in $t_e$ through $R'_e(t_e)$ and is inverted for it.

The scattered pressure generated by the emulated object at $\mathbf{r}_l\in S_l$
then follows as
\begin{subequations} \label{eq:tdpo_shell_all}
\begin{equation}  \label{eq:tdpo_shell}
p_{\mathrm{scat}}(\mathbf{r}_l,t_l)
=\frac{2}{4\pi}\int_{S_{\mathrm{lit}}}
\left[\frac{1}{c}\frac{\partial p_{\mathrm{inc}}}{\partial t'}
+\frac{p_{\mathrm{inc}}}{R'_l}\right]
\frac{\widehat{\mathbf{R}}'_l\cdot\widehat{\mathbf{n}}'}{R'_l}\,
\mathrm{d}s' ,
\end{equation}
the counterpart of \eqref{eq:p_scat_l}, where the factor 2 is that of the PO
surface field, and the bracket is evaluated at the incidence time
$t'=t_l-R'_l/c$. The incident field at the surface point is that of a moving
monopole~\cite{kor2026tdpo},
\begin{equation}  \label{eq:tdpo_p_inc}
p_{\mathrm{inc}}(\mathbf{r}',t')
=\frac{q_e(t_e)}{4\pi R'_e(t_e)\left[1+\mu'_r(t_e)\right]},
\end{equation}
where
$\mu'_r(t_e)=c^{-1}\,\partial_{t_e}\mathbf{R}'_e\cdot\widehat{\mathbf{R}}'_e
=-\mathbf{v}_e\cdot\widehat{\mathbf{R}}'_e/c$ is the directional Mach number
of the emitter toward $\mathbf{r}'$, $\mathbf{v}_e=\mathrm{d}\mathbf{r}_e/\mathrm{d}t_e$
being its velocity, and $1/(1+\mu'_r)$ the Doppler amplitude factor. Since
$\partial t_e/\partial t'=1/(1+\mu'_r)$ by \eqref{eq:time_relations}, the chain
rule gives
\begin{multline}  \label{eq:tdpo_p_inc_dot}
\frac{\partial p_{\mathrm{inc}}}{\partial t'}
=\frac{1}{4\pi R'_e\left(1+\mu'_r\right)^2}\left[\dot{q}_e-\frac{c\,q_e\,\mu'_r}{R'_e}
-\frac{q_e}{1+\mu'_r}\frac{\partial\mu'_r}{\partial t_e}\right],
\end{multline}
\end{subequations}
with $\partial\mu'_r/\partial t_e=c(\mu^2-\mu'^2_r)/R'_e
-\mathbf{a}_e\cdot\widehat{\mathbf{R}}'_e/c$, where $\mu=|\mathbf{v}_e|/c$ is the
Mach number and $\mathbf{a}_e=\mathrm{d}\mathbf{v}_e/\mathrm{d}t_e$ the
acceleration of the emitter. Here $q_e$, $\dot{q}_e$, $R'_e$ and $\mu'_r$ are
evaluated at the emission time $t_e$ associated with $t_l$ through
\eqref{eq:time_relations}, and the integration is again over the illuminated
portion $S_{\mathrm{lit}}$, defined by $\mathbf{R}'_e\cdot\widehat{\mathbf{n}}'<0$
at that emission time.
The motion of the emitter thus enters only through $R'_e(t_e)$, $\mu'_r(t_e)$
and its derivative, evaluated along the prescribed trajectory. For a
stationary emitter, $\mu'_r=0$ and the retarded times in
\eqref{eq:time_relations} become explicit. If the excitation is in addition
harmonic, \eqref{eq:tdpo_shell_all} reduces to \eqref{eq:p_scat_l} under the
substitution $\dot{q}_e \rightarrow \mathrm{j}\omega\widetilde{q}_e$.

The numerical evaluation of \eqref{eq:tdpo_shell_all} must contend with the
implicit relation between the emission and incidence times in
\eqref{eq:time_relations}, handled by the two approaches termed
\emph{backward} and \emph{forward} in~\cite{kor2026tdpo}, with $S_l$ now taking
the role of the observation surface. Only the path from the emitter to the
object depends on the motion: the object and $S_l$ being stationary, the path
from $\mathbf{r}'$ to $\mathbf{r}_l$ contributes the fixed delay $R'_l/c$. With
$S$ discretized into cells centered at $\mathbf{r}'_m$,
\eqref{eq:tdpo_shell_all} is therefore evaluated in two stages. The first
obtains $p_{\mathrm{inc}}$ and $\partial p_{\mathrm{inc}}/\partial t'$ at each
cell on the uniform incidence-time grid $t'_n=n\Delta t$, and the second sums
the contributions of all cells at each control point $\mathbf{r}_l$, delayed by
$R'_{l,m}/c$, with $R'_{l,m}=|\mathbf{r}_l-\mathbf{r}'_m|$.

In stage 1, the backward approach solves
$t_e+R'_{e,m}(t_e)/c-t'_n=0$ for $t_e$ at each cell and grid sample, in closed
form for simple trajectories and numerically otherwise. We adopt the forward
approach, which solves no equation. The emission time is advanced uniformly,
$t_e^{(i)}=i\Delta t$, and each sample is mapped to the incidence time
$t'^{(i)}_m=t_e^{(i)}+R'^{(i)}_{e,m}/c$, with
$R'^{(i)}_{e,m}=|\mathbf{r}'_m-\mathbf{r}_e(t_e^{(i)})|$. At these times,
\eqref{eq:tdpo_p_inc} and \eqref{eq:tdpo_p_inc_dot} are evaluated, or set to
zero if the cell is not illuminated at $t_e^{(i)}$, with $\dot{q}_e$, $\mathbf{v}_e$
and $\mathbf{a}_e$ taken as finite differences of the sampled signal and trajectory.
Successive incidence times are spaced by $(1+\mu'^{(i)}_{r,m})\Delta t$, with
$\mu'^{(i)}_{r,m}=(R'^{(i)}_{e,m}-R'^{(i-1)}_{e,m})/(c\Delta t)$ the discrete
counterpart of $\mu'_r$ toward the cell. The incidence times are thus
compressed where the emitter approaches the cell and dilated where it recedes.
For a subsonic emitter they increase monotonically, so that the cell signals on
the uniform grid $t'_n$ follow by interpolation, provided $\Delta t$ resolves
the shortest period received at the cells.

In stage 2, the delays are applied in the frequency domain, as the phase factors
$e^{-\mathrm{j}\omega R'_{l,m}/c}$ on the discrete Fourier transforms of the
cell signals, so that $p_{\mathrm{scat}}(\mathbf{r}_l,t_n)$ follows without a
second interpolation in time. This uses the whole record, as available in the
present simulations; in real time, the same delays may be applied sample by
sample by interpolation, with fixed fractional delays. Both stages likewise give the pressures on
the offset surfaces $S_l^{\pm}$, and, as in the frequency domain,
$u_{\mathrm{scat}}(\mathbf{r}_l,t_l)=-\partial_{n_l}p_{\mathrm{scat}}$ follows
by a finite difference between them, so that the control signals are now time
series at each $\mathbf{r}_l$ rather than complex scalars.

\begin{figure*}[tb]
    \centering
    \begin{tabular}{c}
        \begin{tabular}{cc}
    \includegraphics[height=4.3 cm, valign = c]{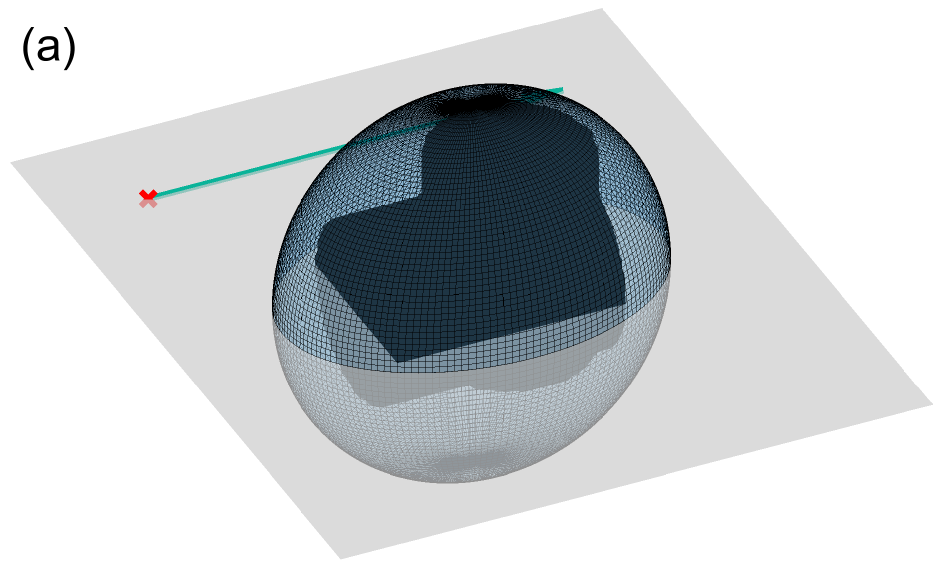}  &  \includegraphics[height=4.3 cm, valign = c]{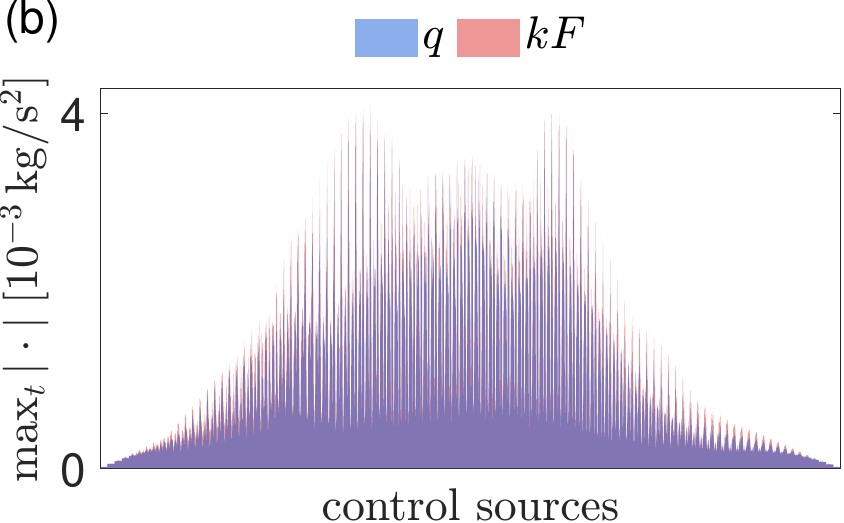}
    \end{tabular} \\
    \includegraphics[width=0.95\linewidth]{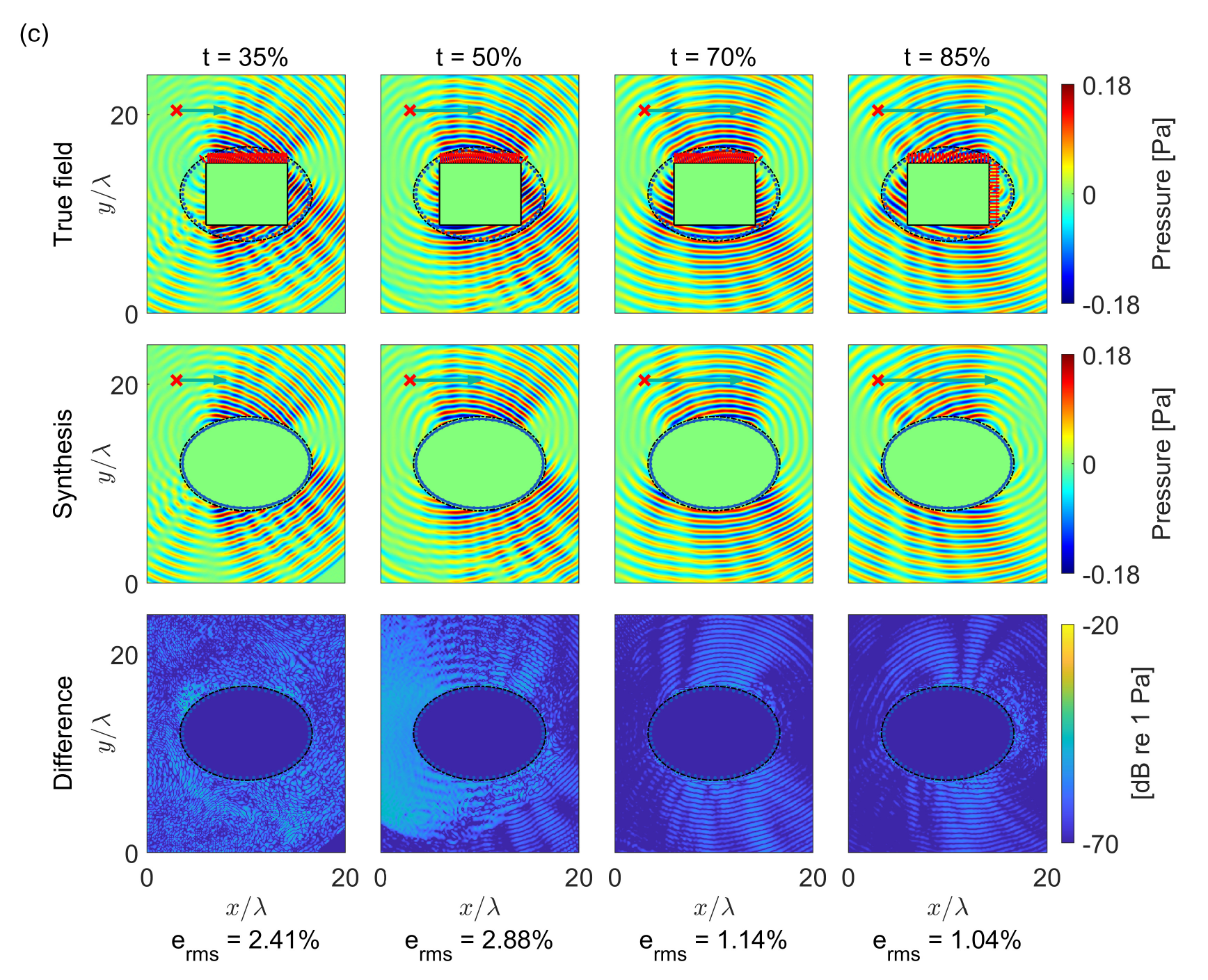}
    \end{tabular}
\caption{Validation of the control algorithm in the time domain via
numerical simulations based on \eqref{eq:td_KH}--\eqref{eq:tdpo_shell_all}.
(a)~Simulation setup, with the emitter trajectory (green arrow) starting at the
red cross. (b)~Peak magnitudes over time of the control signals at all sources.
(c)~Top views of the ROI at four instants, given as fractions of the
trajectory duration: the true scattered field, the synthesized illusion field
and their difference, with the rms error below each column. The black rectangle
is the cross section of the object, with the outward normals at its illuminated
points in red, and the blue curve is the intersection of $S_l$ with the ROI; the
field is not evaluated inside them and is shown as zero.}
    \label{fig:time_sim}
\end{figure*}

The time-domain algorithm is demonstrated with the emulated object, control
surface and emitter signal of Fig.~\ref{fig:freq_sim}, the emitter now moving
along a trajectory in the ROI plane at Mach $0.2$, as shown in
Fig.~\ref{fig:time_sim}(a). The control signals
are evaluated from \eqref{eq:tdpo_shell_all} by the forward approach, and the
reconstruction follows from \eqref{eq:td_KH}--\eqref{eq:td_kernels}. Their peak
magnitudes over the trajectory, $\max_t|q^{(i)}|$ and
$\max_t|kF^{(i)}|$, are shown in Fig.~\ref{fig:time_sim}(b), remaining about
three orders of magnitude below the
emitter amplitude, as in the harmonic case of Fig.~\ref{fig:freq_sim}(f).

Figure~\ref{fig:time_sim}(c) compares snapshots of the reference field, obtained
by direct TDPO evaluation for the physical object, with the synthesized
field and their difference. As the emitter advances, the scattered pattern
sweeps around the object, and the reconstruction follows it closely, with rms
errors of 1.0--2.9\%.

Control sources on an enclosing equivalent surface thus generate the illusion
of a rigid object illuminated by a moving, possibly transient emitter. The
enclosure requirement is relaxed next.

\section{Illusion via an internal control surface}  \label{sec:redatumed}

An enclosing control surface must be at least as large as the object itself,
which becomes impractical as the object grows. We thus consider a control
surface $S_o$ that does not enclose the object, but is instead interior to it
or covers it only partially.

The scattered
field exists only outside the object, so it can be evaluated on $S_l$ but not
on $S_o$, and PO provides no continuation of it into the object. We therefore map the control signals on $S_l$,
$\widetilde{p}_{\textrm{scat}}(\mathbf{r}_l;\omega)$ and
$\widetilde{u}_{\textrm{scat}}(\mathbf{r}_l;\omega)$, onto $S_o$, a contracted
version of $S_l$ meshed at the same spacing, e.g.\ $\lambda/5$, and hence
carrying fewer control points, $L_o<L$. The mapping is formulated in the
frequency domain in Sec.~\ref{sec:redatumed_freq}, and in the time domain,
where causality must be addressed, in Sec.~\ref{sec:redatumed_time}.

\subsection{Frequency-domain formulation}  \label{sec:redatumed_freq}

We again employ the KH representation in
\eqref{eq:holography_S_l}, but now in a reverse (redatuming) sense, expressing
the field on $S_l$ in terms of equivalent source distributions defined on the
internal surface $S_o$. In this context, points $\mathbf{r}_l$ on $S_l$,
previously associated with control sources, are treated as observation points.

The source densities on $S_o$ are written
$\widetilde{p}^{\,\textrm{eq}}_{\textrm{scat}}(\mathbf{r}_o;\omega)$ and
$\widetilde{u}^{\,\textrm{eq}}_{\textrm{scat}}(\mathbf{r}_o;\omega)$, in
analogy with the pressure and scaled normal velocity driving the dipole and
monopole layers on $S_l$, with $\widehat{\mathbf{n}}_o$ the outward unit normal
to $S_o$ at $\mathbf{r}_o$. As on $S_l$, these are densities, in
$\mathrm{kg/(m^2\,s^2)}$ and $\mathrm{Pa}$, related to the lumped strengths of
the redatumed array as in Sec.~\ref{sec:freq}. The superscript
marks them as dummy quantities: no
scattered field exists on $S_o$, and they are defined only by the requirement
that they reproduce the control signals on $S_l$,
\begin{subequations}  \label{eq:holography_redatuming}
\begin{align}
{\widetilde{p}}_{\textrm{scat}}(\mathbf{r}_l;\omega)
&=\int_{S_o}\left(G^{p|q}\,\widetilde{u}^{\,\textrm{eq}}_{\textrm{scat}}
+G^{p|F}\,\widetilde{p}^{\,\textrm{eq}}_{\textrm{scat}}\right)\mathrm{d}s_o,\\
{\widetilde{u}}_{\textrm{scat}}(\mathbf{r}_l;\omega)
&=\int_{S_o}\left(G^{u|q}\,\widetilde{u}^{\,\textrm{eq}}_{\textrm{scat}}
+G^{u|F}\,\widetilde{p}^{\,\textrm{eq}}_{\textrm{scat}}\right)\mathrm{d}s_o,
\end{align}
\end{subequations}
where the arguments $(\mathbf{r}_o;\omega)$ of the source densities and
$(R_{lo};\omega)$ of the kernels are suppressed, with
$\mathbf{R}_{lo}=\mathbf{r}_l-\mathbf{r}_o$ and $R_{lo}=|\mathbf{R}_{lo}|$. 
The kernels $G^{u|q}$ and $G^{u|F}$ give the scaled
normal velocity at $\mathbf{r}_l$ radiated by a monopole and by a dipole at
$\mathbf{r}_o$, and follow from $G^{p|q}$ and $G^{p|F}$ through
$\widetilde{u}_{\textrm{scat}}=-\partial_{n_l}\widetilde{p}_{\textrm{scat}}$.
Explicitly,
\begin{subequations}  \label{eq:G_lo_kernels}
\begin{align}
G^{p|q}(R_{lo};\omega)&=G(R_{lo};\omega),\\
G^{p|F}(R_{lo};\omega)&=G(R_{lo};\omega)\,\alpha_{lo}\,
\widehat{\mathbf{R}}_{lo}\cdot\widehat{\mathbf{n}}_o,\\
G^{u|q}(R_{lo};\omega)&=G(R_{lo};\omega)\,\alpha_{lo}\,
\widehat{\mathbf{R}}_{lo}\cdot\widehat{\mathbf{n}}_l,\\
G^{u|F}(R_{lo};\omega)&=-G(R_{lo};\omega)\left(\beta\eta
+\frac{\alpha_{lo}\,\widehat{\mathbf{n}}_o\cdot\widehat{\mathbf{n}}_l}{R_{lo}}\right),
\end{align}
\end{subequations}
where $\widehat{\mathbf{R}}_{lo}=\mathbf{R}_{lo}/R_{lo}$,
$\alpha_{lo}=\mathrm{j}k+R_{lo}^{-1}$,
$\eta=(\widehat{\mathbf{R}}_{lo}\cdot\widehat{\mathbf{n}}_o)
(\widehat{\mathbf{R}}_{lo}\cdot\widehat{\mathbf{n}}_l)$, and
$\beta=k^2-3\mathrm{j}kR^{-1}_{lo}-3R_{lo}^{-2}$.

The left-hand sides of \eqref{eq:holography_redatuming} are known, having been
obtained from the scattering of the emulated object, so the unknowns are the
two densities on $S_o$. Discretizing the surface integrals as sums over the
$L$ points of $S_l$ and the $L_o$ points of $S_o$ gives the block system
\begin{equation}  \label{eq:Ax_b}
\begin{bmatrix}
\widetilde{\mathbf{p}}_{l} \\
\widetilde{\mathbf{u}}_{l}
\end{bmatrix}
=
\begin{bmatrix}
\mathbf{G}^{p|q} & \mathbf{G}^{p|F} \\
\mathbf{G}^{u|q} & \mathbf{G}^{u|F}
\end{bmatrix}
\begin{bmatrix}
\widetilde{\mathbf{u}}^{\,\textrm{eq}}_o \\
\widetilde{\mathbf{p}}^{\,\textrm{eq}}_o
\end{bmatrix},
\end{equation}
where $\widetilde{\mathbf{p}}_{l}$ and $\widetilde{\mathbf{u}}_{l}$ collect the
data at the $L$ points $\mathbf{r}^{(i)}_l$ of $S_l$, and
$\widetilde{\mathbf{u}}^{\,\textrm{eq}}_{o}$ and
$\widetilde{\mathbf{p}}^{\,\textrm{eq}}_{o}$ the densities at the $L_o$ points
$\mathbf{r}^{(m)}_o$ of $S_o$. Each block
$\mathbf{G}^{\gamma|\delta}\in\mathbb{C}^{L\times L_o}$, $\gamma\in\{p,u\}$,
$\delta\in\{q,F\}$, has entries
$[\mathbf{G}^{\gamma|\delta}]_{i,m}=G^{\gamma|\delta}(R^{(i,m)}_{lo};\omega)$.
In the solution, the
second block row of \eqref{eq:Ax_b} is divided by $\mathrm{j}\omega\rho$, so
that the normal velocity $\widetilde{v}_{\textrm{scat}}$ on $S_l$ is fitted
rather than $\widetilde{u}_{\textrm{scat}}$. The kernels are otherwise
unchanged and the unknowns remain the densities on $S_o$; the scaling only
sets the relative weight of the velocity and pressure data in the fit.

We write \eqref{eq:Ax_b} as $\mathbf{b}=\mathbf{G}_{lo}\mathbf{x}$, with
$\mathbf{b}$ the data vector, $\mathbf{G}_{lo}$ the block matrix and
$\mathbf{x}$ the vector of unknowns. The system is ill-conditioned: propagation from $S_o$ to the
enclosing $S_l$ attenuates the high spatial-frequency content of the densities,
so the inverse amplifies it, together with any error in the data. Moreover, the
monopole and dipole columns of $\mathbf{G}_{lo}$ differ widely in magnitude. We
therefore solve in the general-form Tikhonov sense,
\begin{equation}  \label{eq:tikhonov}
\mathbf{x}_\varepsilon=\left(\mathbf{G}_{lo}^{\dagger}\mathbf{G}_{lo}
+\varepsilon^2\mathbf{D}^2\right)^{-1}\mathbf{G}_{lo}^{\dagger}\,\mathbf{b},
\end{equation}
where $\dagger$ denotes the Hermitian transpose, $\mathbf{D}$ is the
diagonal matrix of the column norms of $\mathbf{G}_{lo}$, and $\varepsilon>0$
is the regularization parameter. This is equivalent to scaling every column of
$\mathbf{G}_{lo}$ to unit norm and applying standard Tikhonov regularization to
the scaled unknowns, so that $\varepsilon$ is relative and each unknown is
penalized in proportion to its influence on the data. Rather than through the
normal equations, which would square the condition number,
\eqref{eq:tikhonov} is computed by QR factorization of the equivalent
augmented least-squares system.
The illusion pressure field at a point $\mathbf{r}$ in the ROI then follows by
the same holography procedure as in \eqref{eq:holography_S_l}, now applied on
$S_o$,
\begin{equation}  \label{eq:holography_S_o}
{\widetilde{p}}_{\textrm{scat}}(\mathbf{r};\omega)
=\int_{S_o}\left(G^{p|q}\,\widetilde{u}^{\,\textrm{eq}}_{\textrm{scat}}
+G^{p|F}\,\widetilde{p}^{\,\textrm{eq}}_{\textrm{scat}}\right)\mathrm{d}s_o,
\end{equation}
where the kernels are now evaluated at $R_o=|\mathbf{R}_o|$ with
$\mathbf{R}_o=\mathbf{r}-\mathbf{r}_o$.

We repeat the frequency-domain simulation of Sec.~\ref{sec:freq} for
the same object and stationary emitter, now driving the contracted surface
$S_o$ rather than $S_l$. The control signals on $S_l$ are those of
Fig.~\ref{fig:freq_sim}, redatumed onto $S_o$ by \eqref{eq:tikhonov}
with $\varepsilon=10^{-3}$, $S_o$ being $S_l$ contracted by a factor of $0.6$.

\begin{figure*}
\centering
\begin{tabular}{c}
\begin{tabular}{cc}
\includegraphics[height = 4.3 cm, valign = c]{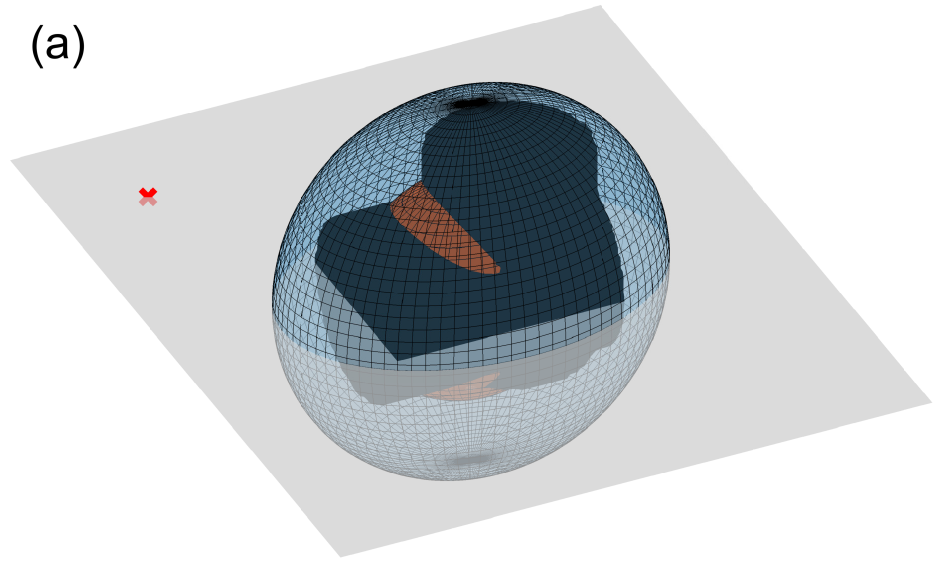} &
\includegraphics[height=4.3 cm, valign = c]{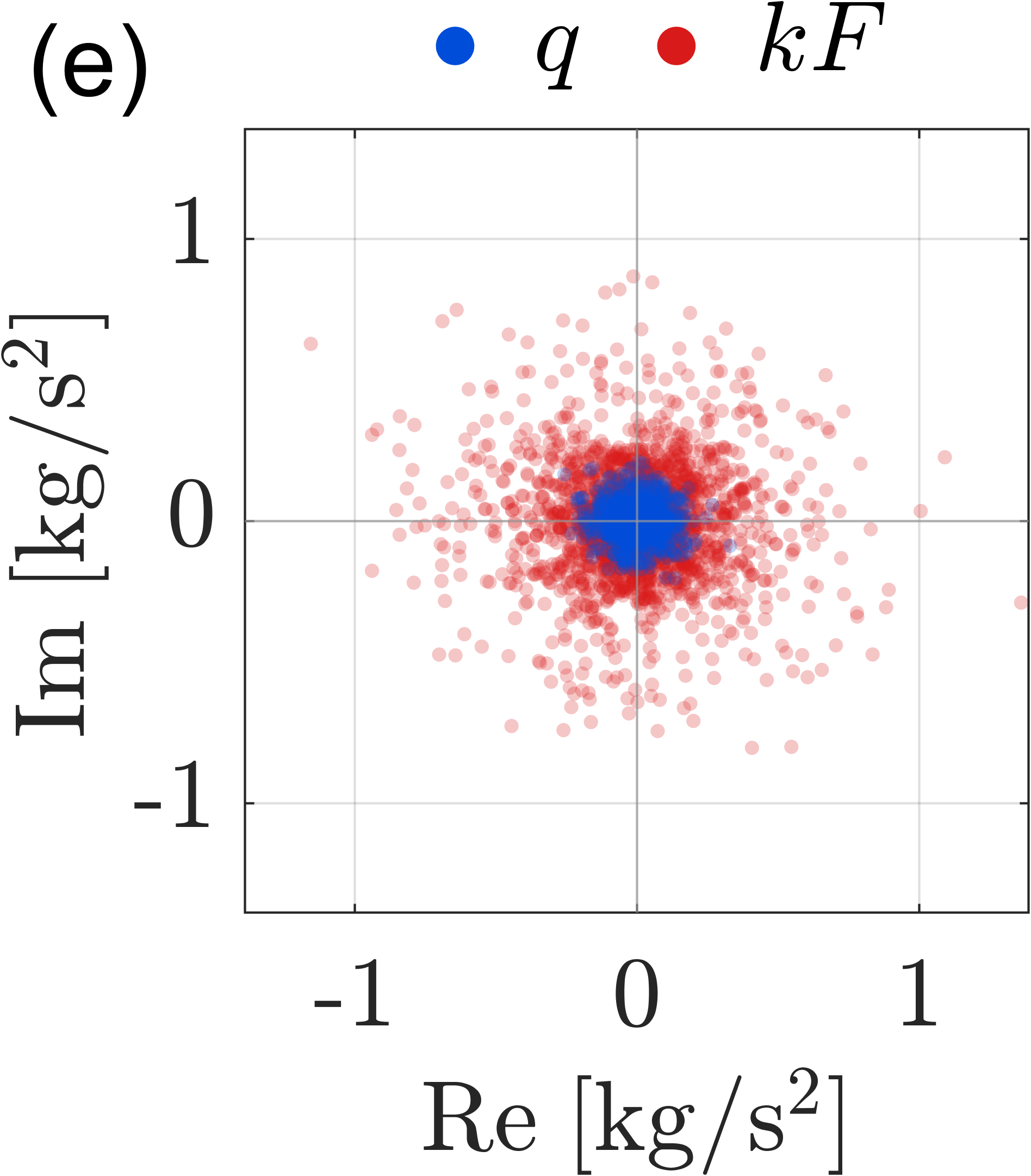}
\end{tabular} \\ \\
\setlength{\tabcolsep}{0pt}
\begin{tabular}{ccc}
\includegraphics[height = 5.1 cm, valign = c]{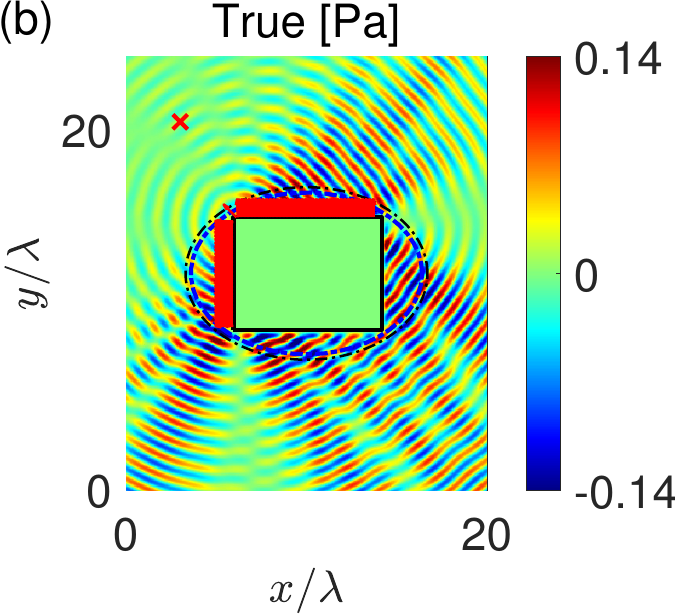}
 &
\includegraphics[height = 5.1 cm, valign = c]{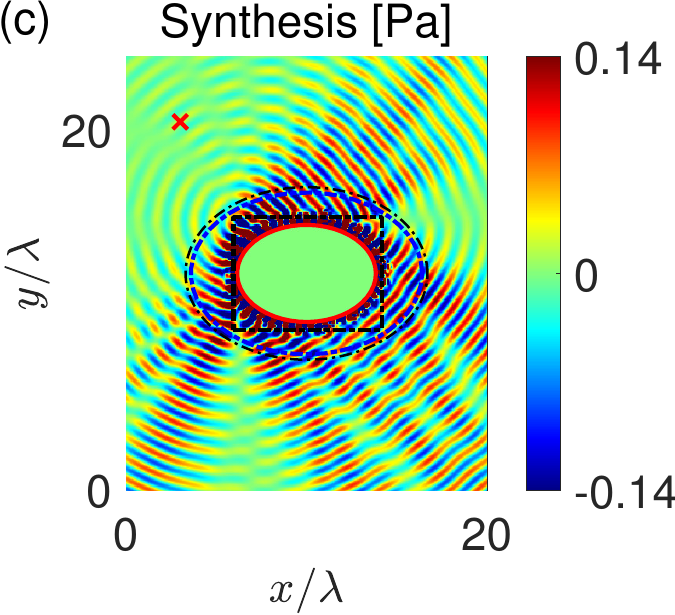} &
\includegraphics[height = 5.1 cm, valign = c]{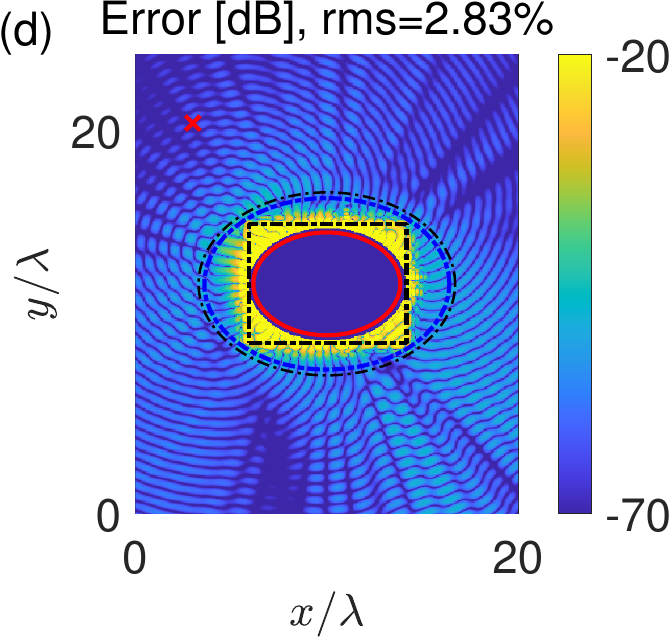}
\end{tabular}
\end{tabular}
\caption{Illusion via an internal control surface, frequency domain.
(a)~Scene: the internal control surface $S_o$ (red) carrying the
redatumed control sources, the enclosing surface $S_l$ (blue) shown for
reference, and the stationary emitter (red cross). (b)~Scattered pressure field
of the physical object, (c)~field synthesized by the sources on $S_o$, and
(d)~their difference. The error is assessed outside $S_l$, the
region between $S_o$ and $S_l$ being excluded. (e)~The control signals at all
sources on $S_o$ in the complex plane.}
\label{fig:redatumed_results}
\end{figure*}

The scene is shown in Fig.~\ref{fig:redatumed_results}(a), with the enclosing
surface retained for reference. Panels (b)--(d) show the true, synthesized and
difference fields, the synthesized field being evaluated from
\eqref{eq:holography_S_o}, and panel (e) the control signals on $S_o$.

Agreement is assessed outside $S_l$ only, since the redatumed densities
are fitted to reproduce the control signals on $S_l$ and the representation
carries no accuracy guarantee in the region between the two surfaces. Outside
$S_l$ the rms error is $2.8$\%, against $1.1$\% with the enclosing surface
of Sec.~\ref{sec:freq}, the increase reflecting the regularized inversion.

\subsection{Time-domain formulation}  \label{sec:redatumed_time}

For a stationary emitter radiating a harmonic signal, the time-domain control
signals follow directly from the densities obtained in
Sec.~\ref{sec:redatumed_freq}. With the convention of
Sec.~\ref{sec:freq}, the sources on $S_o$ are driven by
\begin{subequations}  \label{eq:redatumed_td_signals}
\begin{align}
p^{\,\textrm{eq}}_{\textrm{scat}}(\mathbf{r}_o,t)
&=\left|\widetilde{p}^{\,\textrm{eq}}_{\textrm{scat}}(\mathbf{r}_o;\omega)\right|
\cos\!\left[\omega t+\arg\widetilde{p}^{\,\textrm{eq}}_{\textrm{scat}}(\mathbf{r}_o;\omega)\right],\\
u^{\,\textrm{eq}}_{\textrm{scat}}(\mathbf{r}_o,t)
&=\left|\widetilde{u}^{\,\textrm{eq}}_{\textrm{scat}}(\mathbf{r}_o;\omega)\right|
\cos\!\left[\omega t+\arg\widetilde{u}^{\,\textrm{eq}}_{\textrm{scat}}(\mathbf{r}_o;\omega)\right],
\end{align}
\end{subequations}
each point of $S_o$ carrying a sinusoid of the same frequency, with
amplitude and phase set by the redatumed solution.

Causality now enters. On the enclosing $S_l$ the control signals coincide with
the scattered field and are causal by construction. The sources on $S_o$ lie
inside $S_l$, so their radiation must leave $S_o$ ahead of the field due on
$S_l$, by $R_{lo}/c$ for a pair of points and by at most
$T_o=\max_{\mathbf{r}_l\in S_l,\,\mathbf{r}_o\in S_o}R_{lo}/c$. Each source on
$S_o$ is therefore switched on at the latest instant that still reaches all of
$S_l$ in time, $t_o^{\mathrm{on}}=\min_{\mathbf{r}_l\in
S_l}(t_l^{\mathrm{a}}-R_{lo}/c)$, with $t_l^{\mathrm{a}}$ the first arrival of
the scattered field at $\mathbf{r}_l$. Where this would precede the emission,
the source starts at the emission instant and reaches part of $S_l$ late.

To limit the switch-on transient, each source is also faded in over two periods
by the raised-cosine window
$w(t)=\{1-\cos[\pi(t-t_o^{\mathrm{on}})/T_r]\}/2$ for
$0\le t-t_o^{\mathrm{on}}\le T_r=4\pi/\omega$, with $w=0$ before and $w=1$
after. In the harmonic steady state neither measure costs accuracy: once every
source is fully on, the synthesized field is the same as without them, the
advance being contained in the phases of \eqref{eq:redatumed_td_signals}.

\begin{figure*}
\begin{tabular}{@{}c@{}}
\includegraphics[width=\textwidth]{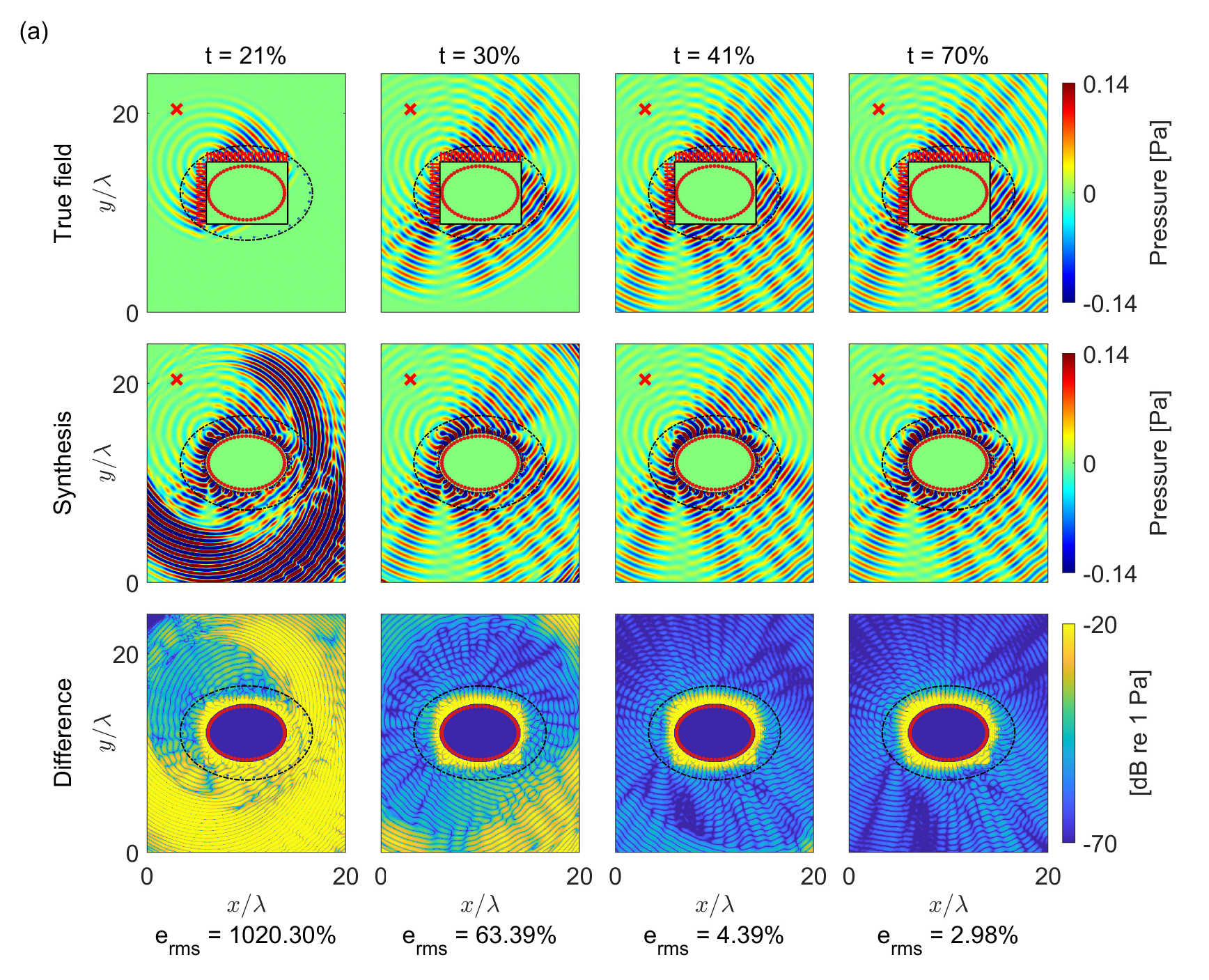}\\[2ex]
\begin{tabular}{cc}
\includegraphics[height = 4.3 cm, valign = c]{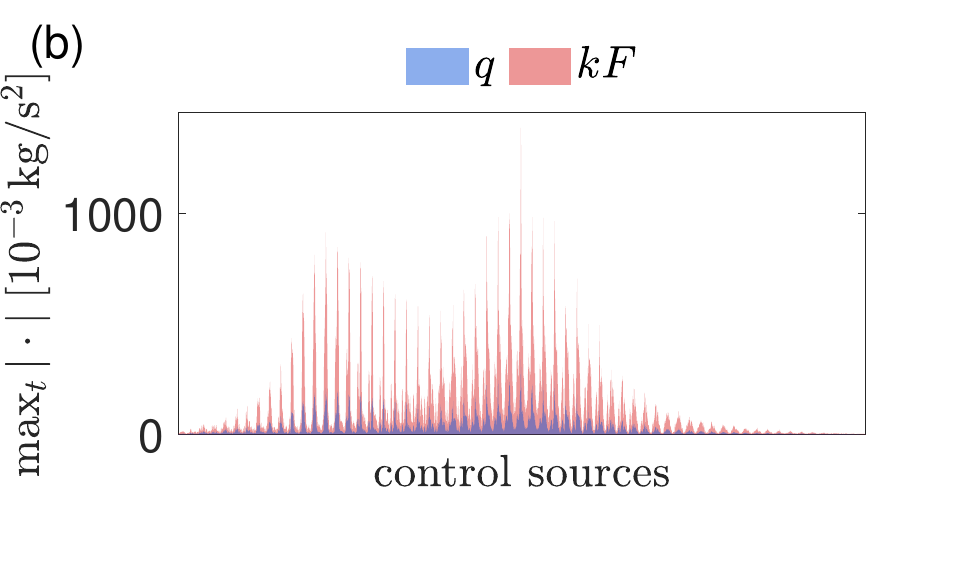} &
\includegraphics[height = 4.3 cm, valign = c]{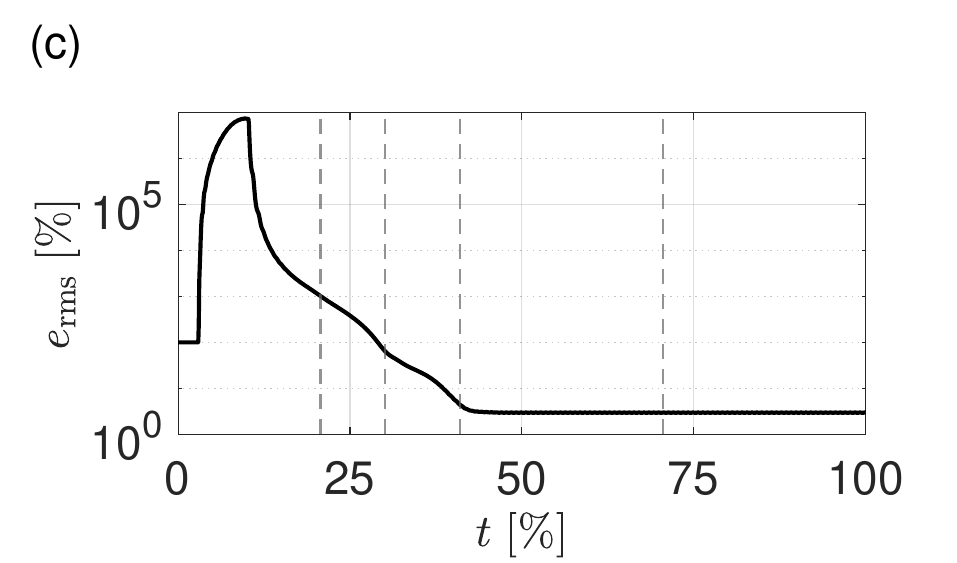}
\end{tabular}
\end{tabular}
\caption{Time-domain illusion via an internal control surface, for the
stationary emitter of Fig.~\ref{fig:redatumed_results}, radiating at
$f=6.88$~kHz ($\lambda=5$~cm), with the sources on $S_o$ spaced by $\lambda/5$,
as on $S_l$. (a)~Top views of the ROI at four instants chosen across the
start-up, given as fractions of the simulation duration of $10.2$~ms, with the rms error below each column. $S_o$ is shown in
red, $S_l$ dashed and the emitter by the red cross; the error is assessed
outside $S_l$. (b)~Peak magnitudes over time of the control signals at all sources.
(c)~Evolution of the rms error, the dashed lines marking the instants of (a).}
\label{fig:redatumed_time_results}
\end{figure*}

The densities on $S_o$ are obtained as in Sec.~\ref{sec:redatumed_freq}, with
the same $\varepsilon$, from the complex amplitudes of the TDPO signals on
$S_l$, extracted by a harmonic fit over their last three periods, and are
converted to the sinusoids \eqref{eq:redatumed_td_signals}. The results, for
the stationary emitter of Fig.~\ref{fig:redatumed_results}, are shown in
Fig.~\ref{fig:redatumed_time_results}. Here $T_o\approx1.5$~ms, and the
switch-on instants range from the emission instant to $0.71$~ms.
The peak control signals are shown in Fig.~\ref{fig:redatumed_time_results}(b).
Figure~\ref{fig:redatumed_time_results}(a) compares snapshots of the reference,
synthesized and difference fields, the error being assessed outside $S_l$ as in
Sec.~\ref{sec:time_domain_regular}. At $21\%$ of the duration the synthesized field is still
building up and departs strongly from the reference; by $41\%$ the error has
dropped to $4.4\%$, and it settles at about $3\%$ once all sources are fully on
and the initial transient has passed, Fig.~\ref{fig:redatumed_time_results}(c).

\section{Conclusion}  \label{sec:conclusion}

In this work, we created a three-dimensional acoustic illusion of a rigid object
illuminated by a moving emitter, using control sources on an equivalent surface.
The sources, monopoles and dipoles, were driven by the field the object would
scatter onto the surface, their signals following in closed form from a
time-dependent physical optics formulation that we developed earlier for a
moving emitter~\cite{kor2026tdpo} and recapped here in \eqref{eq:time_relations}-\eqref{eq:tdpo_shell_all}.

Because the object surface is treated as
locally flat and rigid, the scattered field follows by direct integration over
it, and the motion only shifts the times at which each of its points is
illuminated. The scattering problem is thus not re-solved at each emitter position, nor is the propagation domain
discretized. The control algorithms were tested in simulations on a nonconvex
object composed of two intersecting cylinders. With an enclosing surface, the
rms error was $1.1\%$
for a stationary emitter and $1.0$--$2.9\%$ at Mach $0.2$. The control signals remained
three orders of magnitude below the emitter strength, the effort being
distributed over many sources, whose number is in turn set by the spacing that
a desired level of accuracy requires (here set to $\lambda/5$).

For objects too large to enclose, we proposed a mapping of the control signals
onto a smaller, internal surface that lies inside, or partly inside, the region the object would
occupy and, at the same spacing, carries fewer sources.
The field is then reproduced only outside the enclosing surface, where, for a
stationary emitter, the rms error was $2.8\%$ in the frequency domain and about
$3\%$ in the time domain, of the same order of magnitude as with the enclosing
surface. The penalties are a regularized inversion in place of
the closed form, control signals more than two orders of magnitude larger,
Figs.~\ref{fig:freq_sim}(f) and \ref{fig:redatumed_results}(e), and a start-up
transient wherever causality would require a source to radiate before the
emission. A moving emitter was not considered in this configuration.

Reference and illusion derive from the same PO fields, so these errors measure
the reconstruction, not PO itself. PO is approximate at the rims, and at the
junction of the cylinders, where multiple scattering between them is neglected.

\begin{acknowledgments}
   \textit{
The authors acknowledge the support of the Pazy Foundation, Award number 348072942, and the support of the Israel Innovation Authority, Award number 84993. The authors thank Dirk-Jan van Manen and Amir Boag for insightful discussions.}
\end{acknowledgments}

\bibliography{TDPO}

\end{document}